# Design Of Rubble Analyzer Probe Using ML For Earthquake

Abhishek Sebastian [a)], R Pragna, K Vishal Vythianathan, Dasaraju Sohan Sai, U Shiva Sri Hari Al, R Anirudh and Apurv Choudhary

[1] *Division of Robotics, Abhira A.I, Chennai, Tamilnadu, India.*

[a)] *Corresponding author: abhishek.sebastian2020@vitstudent.ac.in*

**Abstract.** Earthquakes are seismic events that result in the massive destruction of infrastructure, buildings, and human life. Life within each infrastructure is severely affected, and often people are trapped in wrecked building debris. An *earthquake rubble analyzer* is a probe designed to analyze the environment inside rubble and listen to ambient sounds. The designed probe senses human presence by detecting familiar sounds like "hello-help," "breathe," "muffled words," "coughs," and "noise" inside the rubbles using machine learning. The probe is designed to use a TinyML approach and a convolutional neural network (CNN) to classify these sounds with an accuracy of 97.45%. The probe provides real-time sensor data for different parameters like temperature, humidity, air quality, and pressure. The proposed probe can assess the survival prospects of people trapped within the rubble based on factors such as oxygen availability and body fluids retention.

Keywords: Earthquake Rubble, Probe, Arduino, Machine Learning.

## INTRODUCTION

Earthquakes are among the most dangerous and destructive natural disasters [18]. Various events can trigger an earthquake, such as the sudden release of energy in the lithosphere, tectonic plate collisions, or volcanic activity. Earthquakes occur worldwide, although they more frequently hit Japan and Nepal due to their geographical location. Buildings are often damaged, leaving people trapped in the rubble. Those injured under the rubble face a problematic scenario [19]. Workers must first foretell where the survivors are before they can begin digging; thus, this procedure consumes a significant amount of time. Rescuers will have difficulty saving people from the rubble, especially when there is no clear line of sight or too much debris in the path. Temperature, pressure, humidity, and the availability of fresh air are the factors that determine the survivability of a person stuck beneath the rubble [20]. Humans cannot stay in small spaces for a long time, so it is difficult for them to avoid the feeling of claustrophobia [21] when they are in one. Asphyxiation sets in, and body temperature and carbon dioxide levels [22] significantly affect the human body's fluid balance when trapped inside the rubble. The earthquake rubble analyzer probe is a robot designed with 2.4 GHz remote mobility and an Arduino Nano 33 BLE Sense board that can perform tasks involving machine learning. The trained machine learning model recognizes words like "hello" and "help" and noises like cough, breathing, and muffled words inside the rubble. The Arduino Nano 33 BLE Sense board has embedded temperature, pressure, and humidity sensors and an MQ-135 air quality sensor to assess the environmental condition inside the rubble. The ML Model can determine the possibility of the survivor's presence with the rapid deployment of the probe. Experts can monitor the status of survivors and examine how long they can survive, considering the conditions present inside the wreckage. ESP32-CAM module provides Wi-Fi live stream footage from the probe. The probe is a four-wheel-drive robot that provides steady movement during the operation, and its 2.4Ghz Remote Controlling capability allows steady connection over more considerable distances with no discrepancies.

In culmination, this paper contributes to the following.

- A probe designed that can detect people stuck beneath rubble caused due to an earthquake
- A ML Model capable of detecting human sounds using the embedded microphone.
- A system that can provide real-time environmental parameters inside the rubble site using different sensors.
- A system that can provide real-time camera feed of the probe's camera.

# LITERATURE SURVEY

Researchers are working hard to integrate TinyML and low-powered microcontrollers. This section discusses the critical work done in TinyML, Tensor flow lite, Arduino, Different gas sensors, and Human Physiology.

Using Arduino Nano 33 BLE Sense, Alati et al. [1] proposed machine-learning algorithms for greenhouse gas temperature forecasting. The paper focuses on the importance of TinyML applications and Arduino Nano 33 BLE Sense's low-power capabilities. Reddy et al. [2] proposed using an ESP32-CAM board in an IOT-based social distance-checking bot to provide a live stream of the robot's surroundings during operation. Waqar et al. [3] proposed designing a speech anger recognition system using an Arduino Nano 33 BLE Sense's embedded microphone and a custom-trained ML model using TinyML and SER Approach. Anita et al. [4] use Arduino-Mega, Arduino-UNO, Wi-Fi module, and LCD to show the available stopping openings and booking confirmation. Infrared sensors are used at each stop and inform the space accessibility, QR code, if applicable, and data set applications. The Explanatory article on water and electrolyte balance, osmosis, and capillary dynamics has been presented by [6] Watson et al. Farah et al. [7] have proposed using an Arduino UNO interfaced with a gas sensor MQ-135 capable of monitoring air quality along with a SIM900 GSM GPRS shield to send SMS of the sensor data. Ali et al. [8] prepared a research study on the Arduino Nano 33 BLE Sense and its machine learning and other processing capabilities. The following work also emphasizes Arduino Nano 33 BLE Sense's application in wireless sensor networks.

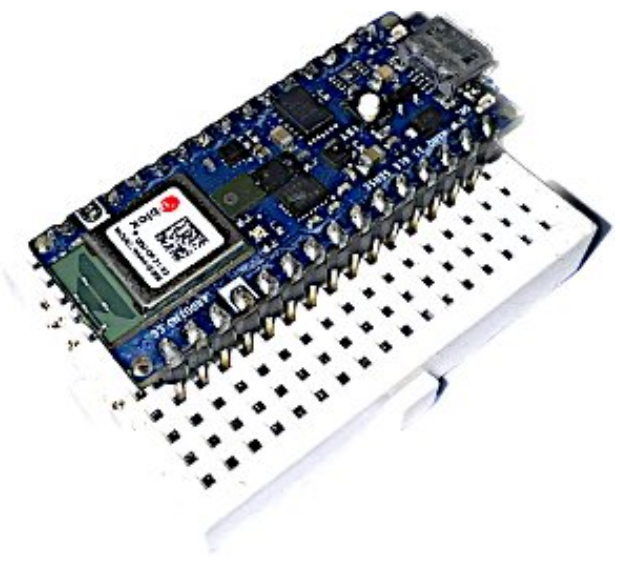

**FIGURE 1.** Arduino Nano 33 BLE Sense

Figure 1 is a pinout diagram of Arduino nano 33 BLE Sense.

In their paper, Kristian et al. [9] discussed Inference speed and quantization of neural networks with TensorFlow lite for Microcontrollers framework. The former's study provides information on the computational capability of Arduino Nano 33 BLE Sense. Shuchang et al. [11] have proposed a low-cost Arduino-based portable device for methane composition in biogas. This system is on an MQ-4 gas sensor and Arduino board. Sai et al.'s study [12] examines the use of the MQ135 sensor to measure air quality and the MQ7 sensor to measure carbon monoxide CO. Kusriyanto et al. [13] have suggested that DHT11, Node-MCU, FC37, and BMP180 sensors can be used to monitor the environment and send readings to a website using an IoT platform. Ruslan et al. [14] proposed the usage of 2.4GHz RC in the development of an Arduino Glove-based self-driving car. Saini et al. [16] proposed an Arduino Uno Board and Zigbee wireless technology design, develop, and test a hardware module that measures meteorological data such as air temperature, dew point temperature, barometric pressure, relative humidity, wind speed, and wind direction. In his book [5], Agus Kurniawan has explained in depth the applications of Arduino Nano 33 BLE Sense through sample hands-on projects. Warden et al. [10] put valuable information on TinyML, Machine Learning with Tensor Flow Lite on Arduino-compatible boards, and other ultra-low power Microcontrollers in the book. Former's book also discusses optimizing energy usage, Optimising the Model, and porting the model from tensor flow to tensor flow lite. Sharan et al. [15] prepared an overview of the current developments in sound recognition technology. Lindasalwa et al. [29] have proposed using a non-parametric method for modeling the auditory perception system, such as Mel Frequency Cepstral Coefficients (MFCCs). Dhingra et al., in their work [30], proposed the approach of isolated speech recognition by using Mel Frequency Cepstral Coefficients (MFCCs) and Dynamic time warping (DTW). Mie Mie Oo [31], in his comparative study of MFCC feature with different ML techniques in acoustic scene classification, has compared the properties of various classifiers such as K-nearest neighbors (KNN), Support Vector Machine (SVM), Decision Tree (ID3) and Linear Discriminant Analysis by using MFCC.

# METHODOLOGY

This section outlines the experimental setup and technique used to assess the designed model's performance.

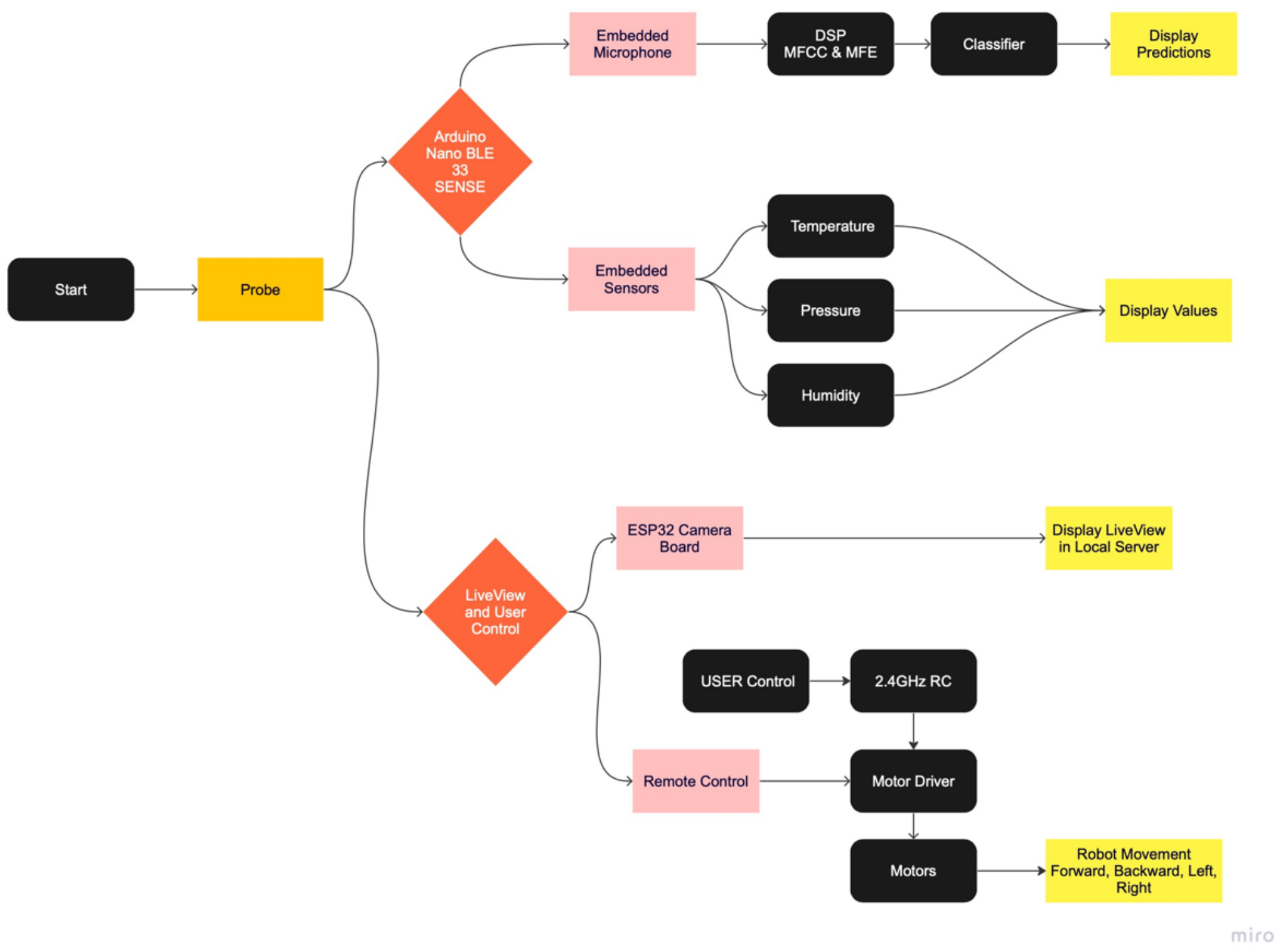

**FIGURE 2.** Working methodology of earthquake rubble analyzerprobe.

The above figure 2 explains the operational flow of the earthquake rubble analyzer probe. The working methodology of the probe mainly comprises 2 Modules:

An ESP 32 camera module is in the designed earthquake rubble analyzer probe. The ESP 32 camera comes with an OV2640 camera and provides an onboard TF card slot. The rubble's camera module helps to observe the area of interest and maneuver the probe through the rubble. The probe body primarily consists of motors, custom designed motor driver, 2.4Ghz Remote control, and Li-Po Battery. The custom motor driver controls the motors, which get power from the Li-Po Battery and User inputs from the 2.4Ghz Remote Control. Arduino Nano BLE 33 Sense is the main CPU used in the rubble analyzer probe; it has embedded sensors to sense the surrounding temperature, humidity, and pressure. The serial monitor displays the obtained sensor data. The Arduino Nano BLE 33 Sense also has an embedded microphone that records the ambient audio for 2 seconds and pre-processes it using DSP algorithms such as MFCC & MFE to obtain the features. The obtained features are then sent to the classifier to detect human presence.

# HARDWARE DEVELOPMENT

The Earthquake rubble analyzer is a four-wheeled robot with a custom-designed motor driver that a 2.4 GHz remote control can control. The probe uses Arduino Nano 33 BLE sense for machine learning and sensor integration. An nRF52840 (Cortex M4F) processor powers this board, which can clock up to 64MHz and has a 1MB flash memory and 256KB SRAM [5]. The robot's body consists of four (5V) DC motors powered by a 12 Volt three-cell Li-Po battery. The custom motor driver provides power and control to the motors. The motor driver also has provisions for interfacing with the 2.4GHz receiver to maneuver the robot using a suitable controller.

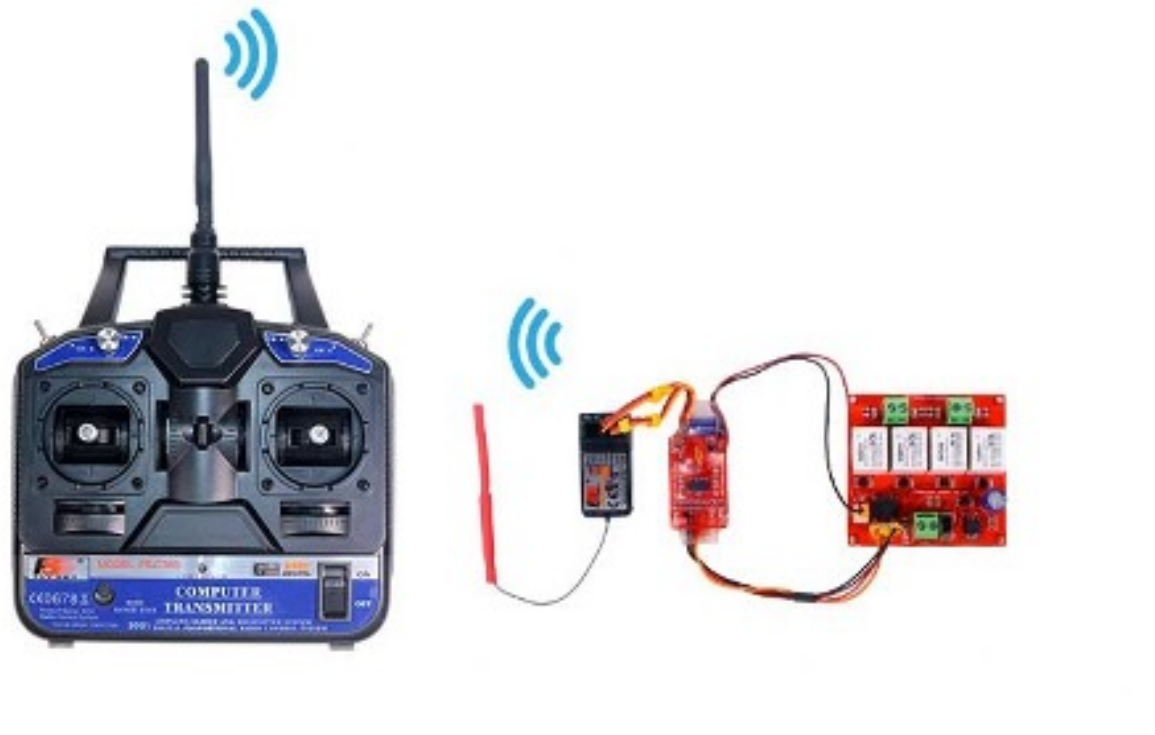

**FIGURE 3.** 2.4Ghz Remote Control with Custom-Motor Driver

Figure 3 is a connection diagram between 2.4GHz remote control and the custom motor driver which is used in the probe.

The sensors used in the probe to analyze the environment are HTS221, LP22HB, and MQ135. The HTS221 is a mixed-signal ASIC-equipped sensor element that is a super-compact relative humidity and temperature sensor, and it transmits measurement data via digital serial ports [5]. The LPS22HB is a piezoresistive absolute pressure sensor with a digital output barometer. The device incorporates a sensing component and an IC interface that uses I2C or SPI to link the component to the application. The MQ-135 gas sensor detects Ammonia ($NH_3$), sulfur (S), benzene ($C_6H_6$), $CO_2$, smoke, and other dangerous gasses. Like the other gas sensors in the MQ series, has both a digital and an analog output pin. When the amount of these gases in the air exceeds a threshold limit, the digital pin goes high [18].

## SOFTWARE DEVELOPMENT

The Proposed Earthquake analyzer probe uses TensorFlow Lite for microcontrollers (TFLM). TensorFlow Lite for microcontrollers is an ML algorithm execution interface [23]. An experimental adaptation of TensorFlow Lite designed for microcontrollers with only a few KB of memory is called TensorFlow Lite for microcontrollers. One of the critical advantages of TFLM is that existing tensor flow environments are used for development, training, and testing [9]. If a microcontroller is the ultimate deployment device, a model must be translated to TensorFlow Lite and subsequently to TFLM data structures. The first step is to get Keras or equivalent framework training. Once the findings are satisfactory, the model can be stored. A model must be converted to TF Lite format after it has been saved.

### Training workflow

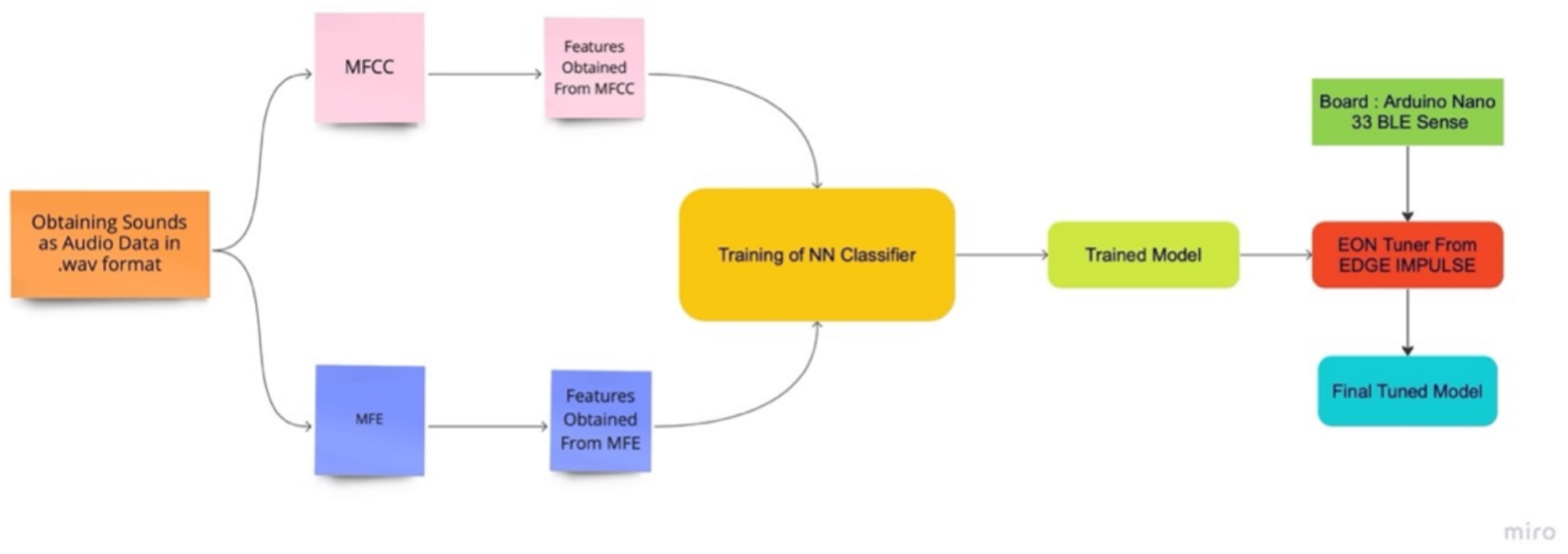

**FIGURE 4.** Model Training Methodology

The above figure 4 explains the working flow of custom model training, to detect sounds of interest using Edge impulse.

Training the model has five significant steps, starting with the preparation of the dataset for all possible scenarios, performing suitable digital signal processing procedures like MFCC (Mel Filter Crystal Coefficients) [24], MFE (Mel Filter bank Energy) [25], and get features for the audio files. A neural network audio classifier is trained based on TensorFlow with processed features and raw data. The model is also tuned using an open-source tool for efficiency and higher computational ability. The tuned model is then tested to substantiate accuracy. The following section briefs the above steps in detail.

## Data set

Data acquisition is the process of acquiring sound data using a microphone. Different sound data is collected within a length of 1 second. The data set is collected for around 134 minutes, with an equal split for five different scenarios. A total of 8040 and 1608 audio files for each scenario are recorded. The training data set and test data set split ratio are set to 84% and 16%, respectively.

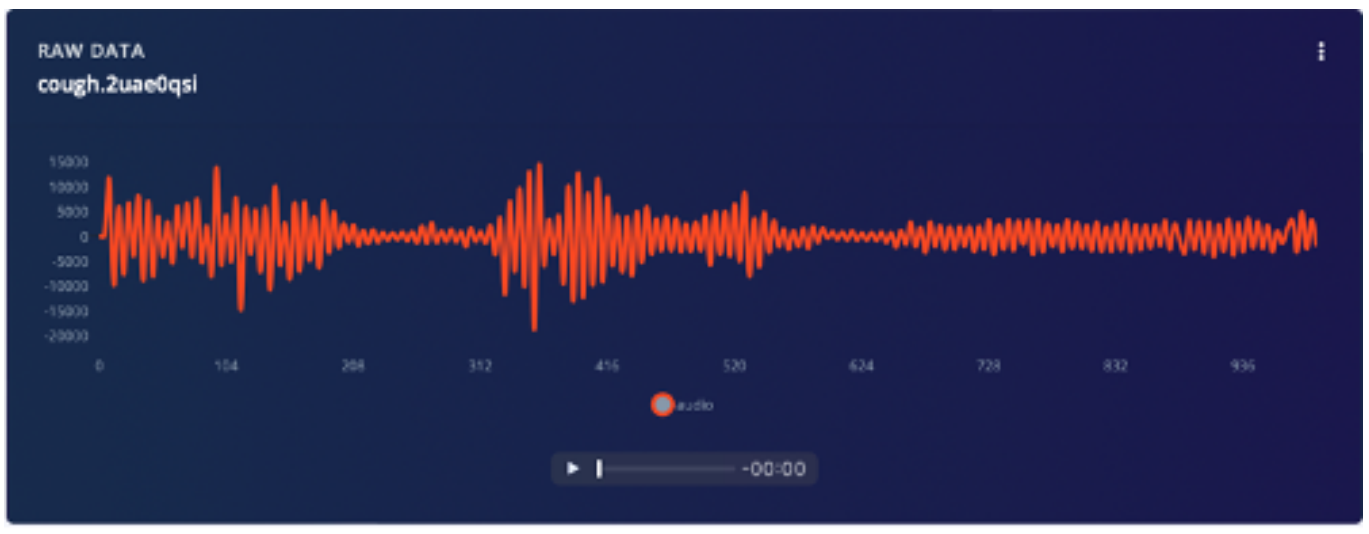

**FIGURE 5.** Excerpt from data acquisition Process

The above figure 5 is an excerpt of a “Cough” Recording from data acquisition process.

## Data pre-processing

MFCC and MFE are performed on the data set to obtain features for the machine learning model. A signal's temporal and frequency properties are extracted by performing MFCC and MFE. The frequency domain employs a non-linear scale known as the Mel scale. It works effectively with audio data, primarily for non-voice recognition applications where the sounds to be categorized are easily distinguishable by the human ear.

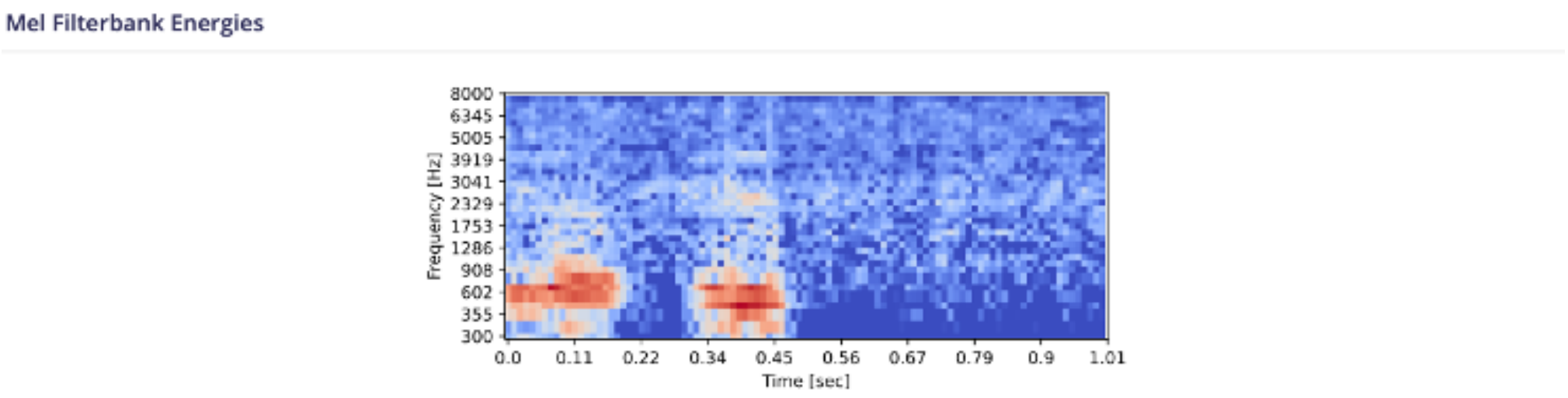

**FIGURE 6.** Excerpt from Data Pre-processing

The above figure 6 is an excerpt of a “Hello-Help” recording’s MFE Spectrogram from data Pre-Processing process.

By performing pre-processing techniques like MFCC and MFE , features from the audio data files are extracted to use the same for model training.

## Machine learning

The neural network classifier used here is Keras [26], which is suitable for training audio signals, the ratio of training set and validation set is 0.8:0.2, with a learning rate of 0.0005 and 100 epochs. The model trained is based on Sequential from TensorFlow. Adam [27] is used as the model optimizer [32].

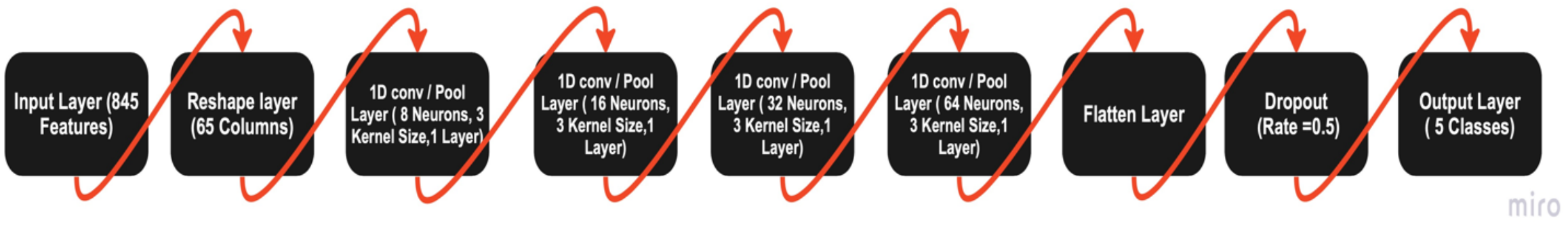

**FIGURE 7.** Model Architecture

The above Figure 7 explains the architectural flow for the trained model.

**TABLE 1.** Confusion Matrix with respect to validation dataset

| | **Breath** | **Cough** | **Hello, Help** | **Muffled Words** | **Noise** |
|---|---|---|---|---|---|
| **Breath** | 75% | 8.3% | 0% | 16.7% | 0% |
| **Cough** | 0% | 87.5% | 0% | 12.5% | 0% |
| **Hello, Help** | 6.3% | 0% | 93.8% | 0% | 0% |
| **Muffled Words** | 7.1% | 14.3% | 7.1% | 64.3% | 7.1% |
| **Noise** | 0% | 0% | 0% | 0% | 100% |
| **F1 Score** | 0.78 | 0.78 | 0.94 | 0.69 | 0.96 |

From the **TABLE 1**, It can be inferred that the accuracy of the model is 83.6% with respect to validation set.

## Eon tuner

The EON tuner is an open-source development tool from **Edge Impulse [28]** that allows developers to choose the optimum trade-offs for their applications. The input data, signal processing elements, and neural network architectures are all analyzed by the EON Tuner. It provides an overview of potential model architectures that will meet the latency and memory requirements of the selected device (Here, an Arduino Nano 33 BLE sense board). This process tunes the neural network to Arduino Nano BLE 33 Sense considering its restrictions and capabilities.

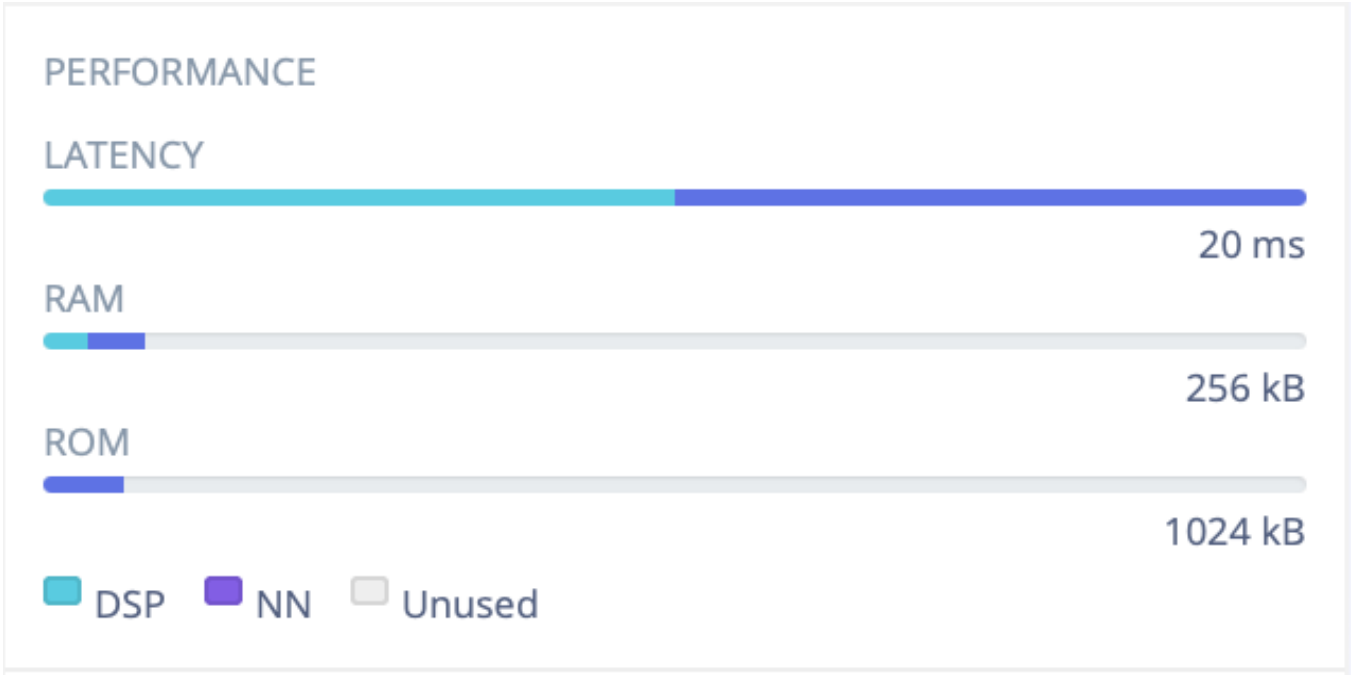

**FIGURE 8.** Performance parameters after EON tuning the model

The above image figure 8 is an excerpt from the performance analysis of the model after EON Tuner. The RAM and ROM usage of the board has been reduced considerably to improve performance.

## Model testing

The model is tested with testing data acquired with an overall accuracy of 89.83%, with individual F1 scores of 0.87 for breath, 0.91 for cough, 1.00 for hello-help, 0.87 for muffled words and 0.95 for noise.

**TABLE 2.** Confusion Matrix of trained model with respect to testing dataset

| | Breath | Cough | Hello, Help | Muffled Words | Noise | Uncertain |
|---|---|---|---|---|---|---|
| **Breath** | 83.3% | 0 % | 0% | 8.3% | 0% | 8.3% |
| **Cough** | 8.3% | 83.3% | 0% | 8.3% | 0% | 0% |
| **Hello, Help** | 0% | 0% | 100% | 0% | 0% | 0% |
| **Muffled Words** | 0% | 0% | 0% | 86.7% | 6.7% | 6.7% |
| **Noise** | 0% | 0% | 0% | 0% | 100% | 0% |
| **F1 Score** | 0.87 | 0.91 | 1.00 | 0.87 | 0.95 | |

From the above **TABLE 2.** it can be inferred that the accuracy of the model is 89.83% with respect to validation set. The improvement in accuracy is influenced by **EON Tuner**. The model's accuracy is mainly affected by muffled words and breath, which is has similar frequency components making it difficult for the model to classify them.

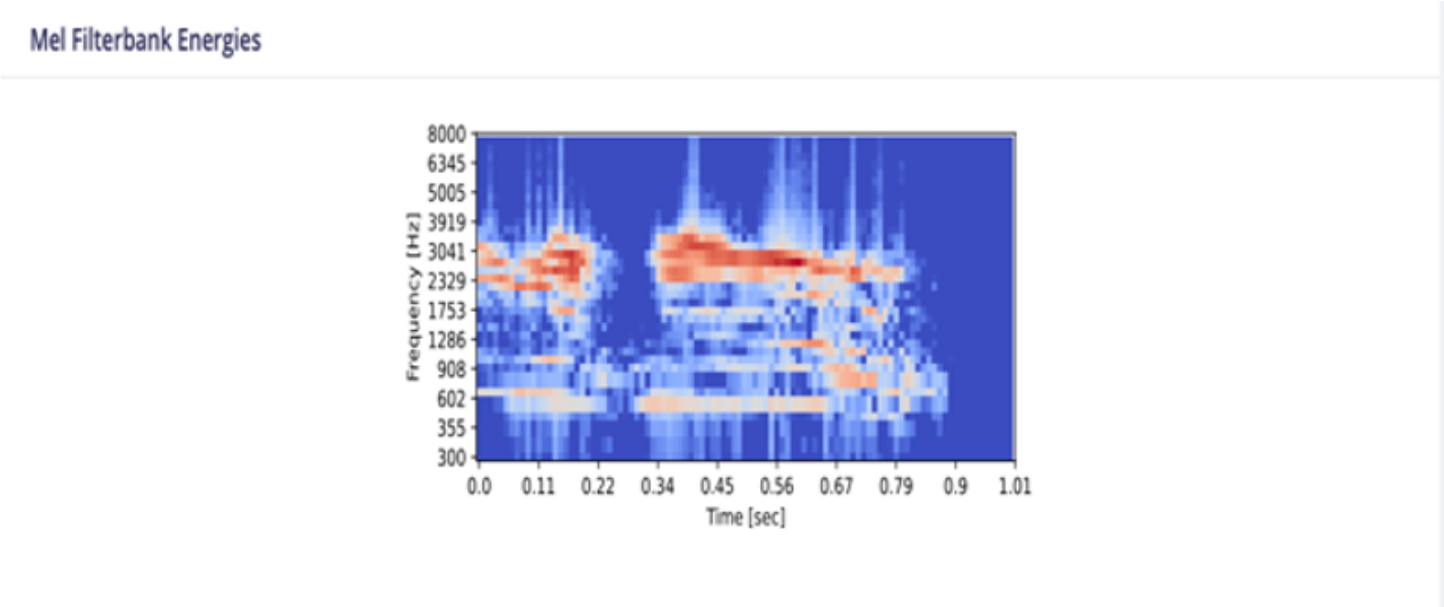

**FIGURE 9.** MFE excerpt of a "Muffled words" audio file.

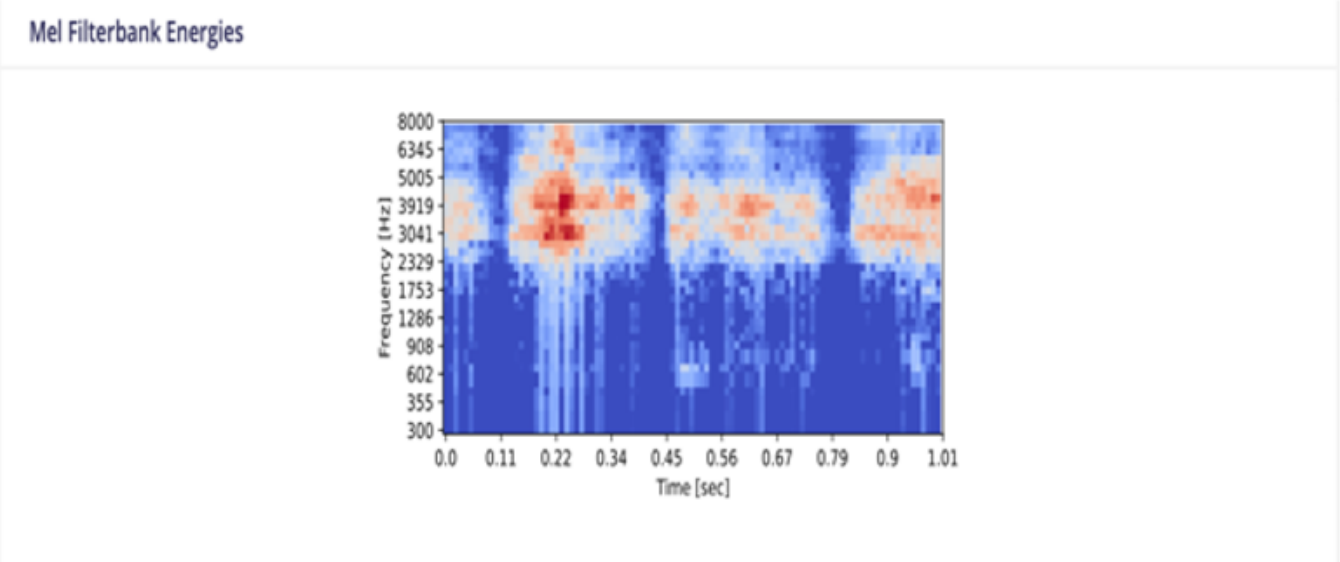

**FIGURE 10.** MFE excerpt of a "breath" audio file.

The above images Figure 9 and Figure 10 show the MFE excerpts of audio files, from which it can be inferred that the major frequency properties lie around the same frequencies (2000 Hz – 3000 Hz).

## Working setup of the probe

Once it is booted up, the probe starts to record valuable data from its surroundings. The recorded sound is then processed and fed into the model to detect any sounds of interest. Once the process is over, the serial monitor connected to the computer displays the predictions. Similarly, the inbuilt embedded sensors (HTS221, LP22HB) and MQ135 then record their respective parameters, Moreover, print them on the serial monitor.

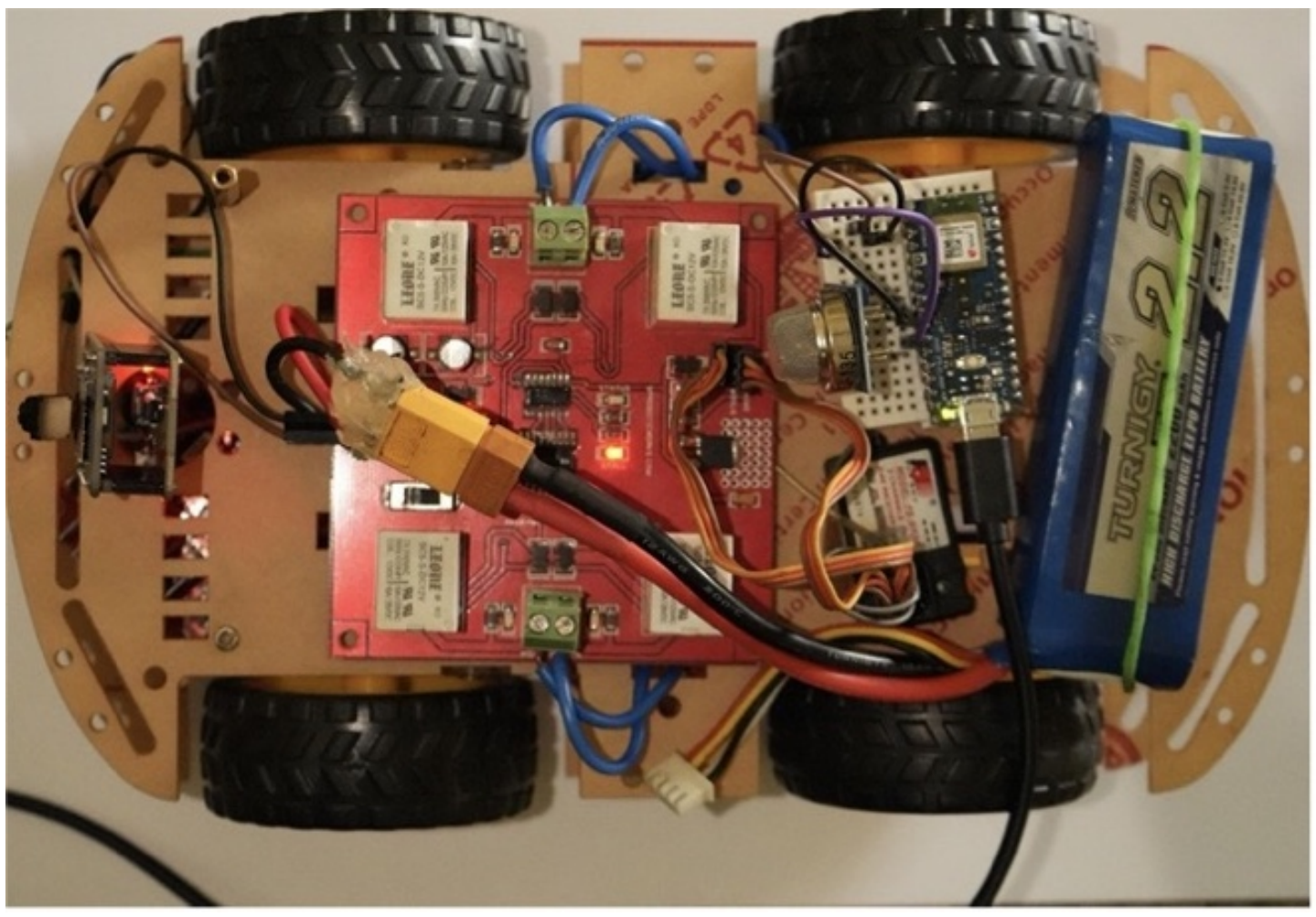

**FIGURE 11.** Top View of the probe

Figure 11 shows the top view of the earthquake rubble analyzer probe.

The ESP32 CAM board connects to the Wi-Fi, sets up a local server, and broadcasts the live feed. Before booting the motor driver, the 2.4GHz switches for seamless pairing-up and connection with the 2.4Ghz receiver. After completing the procedure, the bot is ready to be deployed in the field.

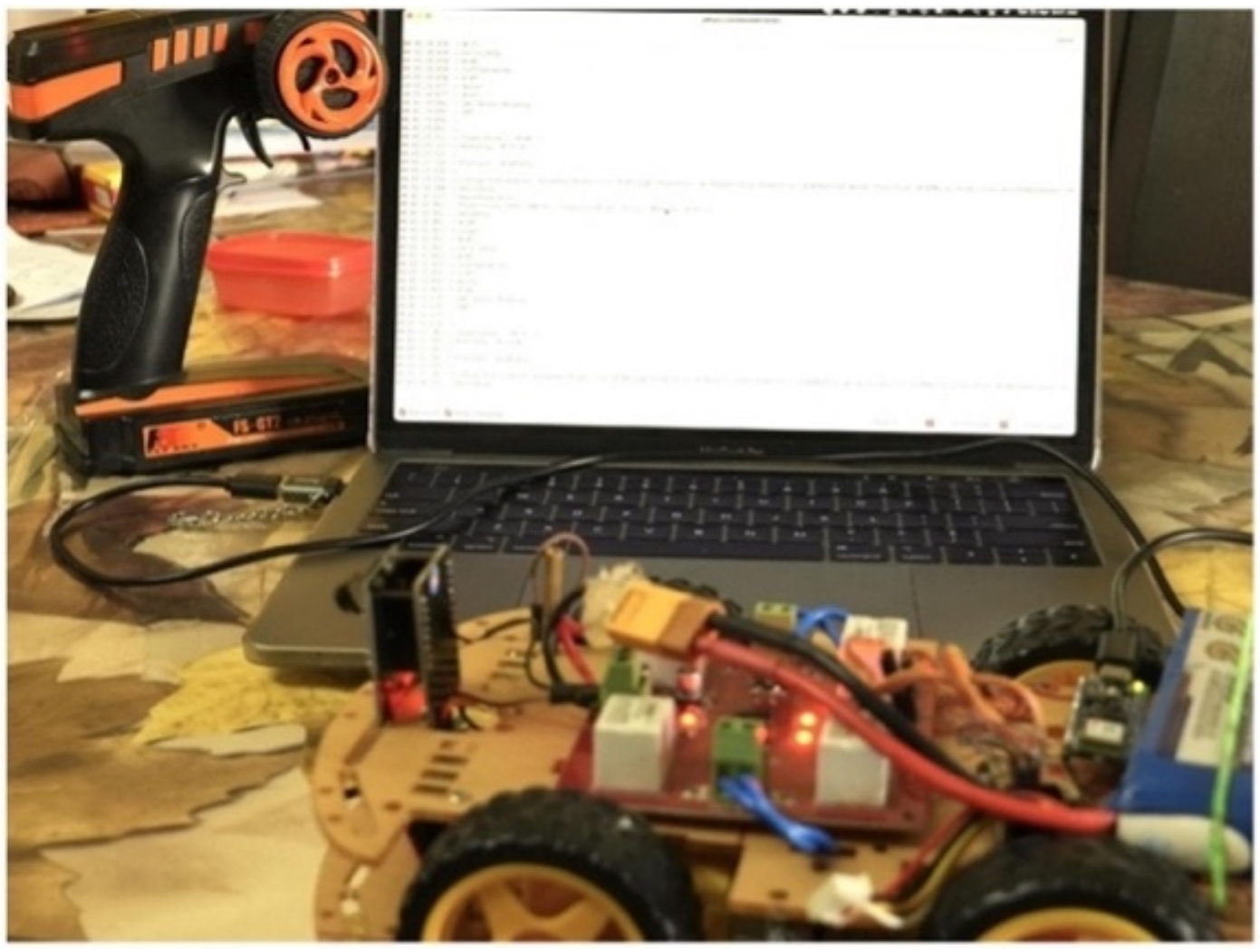

**FIGURE 12**. Work station of the probe

Figure 12 shows the transmitter, the serial monitors and the probe in a ready to deploy state.

The transmitter can control the bot's movement by accessing the local server, where the surroundings are in view. Once it is out of sight of the pilot's view, a long UART cable connects the Arduino Nano BLE 33 Sense. The predictions and sensor readings can be accessed using the Arduino serial monitor.

## EXPERIMENTAL DISSCUSSIONS AND RESULTS

The following section deals with the experimental outcomes and discussions of the same.
The results from the experiments are as follows:

1. Machine learning results
2. Sensor data results

## Machine learning results

The microphone embedded in the Arduino Nano BLE 33 Sense starts to record audio for 1 second from its surroundings and predicts different sounds heard by the microphone.

```
22:56:14.928 -> Starting inferencing in 2 seconds...
22:56:16.683 -> Recording...
22:56:17.658 -> Recording done
22:56:17.976 -> Predictions (DSP: 304 ms., Classification: 19 ms., Anomaly: 0 ms.):
22:56:17.976 -> breathes:
22:56:17.976 -> 0.00
22:56:17.976 -> cough
22:56:17.976 -> 0.07
22:56:17.976 -> hello,help:
22:56:17.976 -> 0.07
22:56:17.976 -> muffled_words:
22:56:17.976 -> 0.65
22:56:17.976 -> Noise:
22:56:17.976 -> 0.21
```

**FIGURE 13.** Excerpt from machine learning results

From the image figure 13, There is a 7% possibility of cough detection, a 65% possibility of muffled noise detection, and the rest 21% of the sound could be noise.

## Sensor data results

The performance of the suggested system is validated based on the following experiments:

- The results of professional-grade medical equipment readings were adequately collated.
- Additionally, the readings from the proposed system were carefully tabulated.
- These readings were taken five times at an interval of 30 minutes each.
- To verify the effectiveness of the suggested system, the two readings were compared.
- The readings tabulated were used to calculate deviation and percentage error.

Different sensor data from sensors are obtained through serial communication and displayed in the serial monitor. The sensor data from the probe validates with a professional-grade digital thermometer, humidity meter, and digital barometer.

**TABLE 3**. Professional grade equipment used to validate proposed probe's sensor readings

| S. No | Name of the Device | Purpose |
|---|---|---|
| 1 | HTC Digital Barometer Altimeter AL-7010 | To measure surrounding pressure. |
| 2 | HTC HD-303 Humidity Temperature Meter. | To measure surrounding temperature and humidity. |

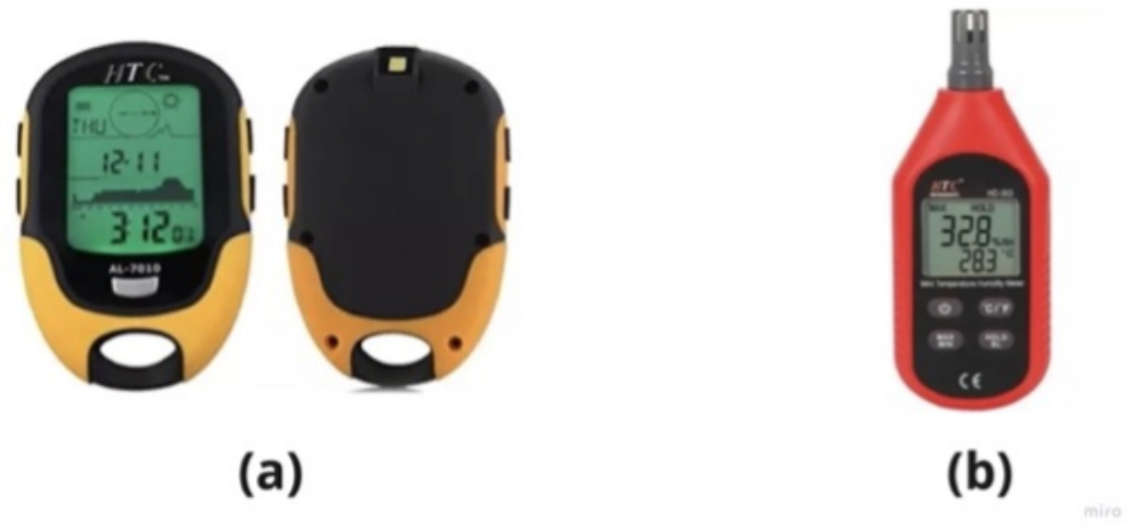

**FIGURE 14.** Professional grade equipment used to validate proposed probe's sensor performance.

Figure 14 (a) shows the HTC digital barometer altimeter AL-7010, and (b) shows the HTC HD-303 humidity temperature meter.

**TABLE 4.** Readings from the professional grade thermometer and proposed probe's temperature sensor values.

| S. No | Readings from the HTC HD-303 Humidity Temperature Meter. (in °C) | Readings from the Proposed probe's temperature sensor values. (in °C) | Deviation (in °C) | Percentage Error |
|---|---|---|---|---|
| **1** | 32.12 | 32.04 | -0.08 | 0.2497% |
| **2** | 30.07 | 30.45 | +0.38 | 1.248% |
| **3** | 31.56 | 31.63 | +0.07 | 0.2213% |
| **4** | 31.67 | 31.78 | +0.11 | 0.3461% |
| **5** | 32.45 | 32.39 | -0.06 | 0.18524% |

From **TABLE 4**, the average percentage error is 0.4590%. In other words, the proposed probe's temperature sensor is 99.540% accurate.

**TABLE 5.** Readings from the professional grade humidity meter and proposed probe's humidity sensor values.

| S. No | Readings from the HTC HD-303 Humidity Temperature Meter. (in %) | Readings from the Proposed probe's temperature sensor values. (in %) | Deviation (in %) | Percentage Error |
|---|---|---|---|---|
| **1** | 52.88 | 52.81 | -0.07 | 0.13238% |
| **2** | 51.67 | 51.45 | -0.22 | 0.4258% |
| **3** | 55.34 | 55.38 | +0.04 | 0.07228% |
| **4** | 59.67 | 59.75 | +0.08 | 0.13407% |
| **5** | 54.34 | 54.46 | +0.12 | 0.22083% |

From **TABLE 5**, the average percentage error is 0.202%. In other words, the proposed probe's humidity sensor is 99.798% accurate

**TABLE 6**. Readings from the professional grade pressure meter and proposed probe's pressure sensor values.

| S. No | Readings from the HTC Digital Barometer Altimeter AL-7010 (in kPa) | Readings from the Proposed probe's Pressure sensor values. (In kPa) | Deviation (In kPa) | Percentage Error |
|---|---|---|---|---|
| **1** | 0.011 | 0.01 | -0.001 | 9.09% |
| **2** | 0.012 | 0.01 | -0.002 | 16.67% |
| **3** | 0.011 | 0.01 | -0.001 | 9.09% |
| **4** | 0.010 | 0.01 | 0 | 0% |
| **5** | 0.010 | 0.01 | 0 | 0% |

From **TABLE 6**, the average percentage error is 6.97%. In other words, the proposed probe's humidity sensor is 93.03% accurate. The collective accuracy from the testing is 97.456% for the three proposed sensors in the probe.

```
22:56:18.224 -> GAS Sensor Reading:
22:56:18.224 -> 168
22:56:18.224 ->
22:56:18.224 ->
22:56:18.224 -> Temperature = 32.67 °C
22:56:18.224 -> Humidity= 52.81 %
22:56:18.224 -> --------------
22:56:18.224 -> Pressure = 0.00 kPa
22:56:18.224 -> --------------
```

**FIGURE 15.** Excerpt from sensor data results

Figure15 shows the sensors data output printed in the serial monitor.

The gas sensor readings range from 0-1024. From the image above, we can infer that the reading is 168, considered healthy air quality.

The probe is the first to analyze earthquake rubble for human life. There are similar works from researchers for different applications like weather stations, gas alert systems, anger prediction from speech, and Arduino-based self-driving cars. Research work done behind this probe can constructively improve possibilities of saving human and animal lives struck under rubble during earthquakes.

## CONCLUSION

It can be concluded from this research that a probe is to analyze the environment inside a rubble site using various sensors. The proposed system can successfully determine the survival rate of the person trapped by analyzing different parameters like temperature, humidity, pressure, and the environment of the rubble. The designed machine learning model predicts the presence of any person trapped beneath the rubble and the survival rate with accepted accuracy. The probe can also detect familiar sounds and report them using machine learning and an embedded microphone. The probe using 2.4GHz can be controlled seamlessly for longer distances without much signal interference. The overall accuracy of the designed Earthquake rubble analyzer probe is 97.45%.

The future iteration robot can be wireless by transmitting predictions and sensor data without a UART cable. The robot's chassis design can be customized efficiently for more straightforward accessibility in confined spaces. The design of the robot can be worked on and modified in the future.